# Observation of a new polyhalide phase in Ag-Cl$_2$ system at high pressure

Adam Grzelak[1]*, Jakub Gawraczyński[1], Mariana Derzsi[2], Viktor Struzhkin[3], Maddury Somayazulu[4], Wojciech Grochala[1]

[1]Center of New Technologies, University of Warsaw, Zwirki i Wigury 93, 02089 Warsaw, Poland
[2]Advanced Technologies Research Institute, Faculty of Materials Science and Technology in Trnava, Slovak University of Technology in Bratislava, J. Bottu 8857/25, 917 24 Trnava, Slovakia
[3]Center for High Pressure Science and Technology Advanced Research, Shanghai 201203, China
[4]HPCAT, X-ray Science Division, Argonne National Laboratory, Lemont, IL 06439, United States

*a.grzelak@cent.uw.edu.pl

**Abstract**

In this short contribution, we examine Raman spectroscopic data from high-pressure and high-temperature experiments with Ag-Cl$_2$ system, and find that they are in good agreement with previously observed and calculated spectra of polychloride species. Our results imply the formation of a hitherto unknown AgCl$_x$ compound, which warrants further study.

**Introduction**

Among the three coinage metals known since antiquity – copper, silver and gold – the chemistry of silver is perhaps the most elusive and challenging [1]. It adopts the oxidation state +1 (with closed-subshell d$^{10}$ electronic configuration) in the vast majority of its known compounds [1]. However, it is the chemistry of silver(II) that has received a much greater attention from solid state chemists and physicists, due to the unpaired d electron giving rise to magnetic interactions in AgF$_2$ and its ternary derivatives [2,3]. Ag$^{2+}$ cation is also one of the strongest known oxidizers and because of that, it is found predominantly in the coordination environment of fluorine (the most electronegative element) as the ligand in known thermodynamically stable compounds of Ag$^{II}$.

In this work, we are concerned with possibility of obtaining novel compounds of silver and chlorine (the third most electronegative element). Phase diagram of binary Ag/Cl system features only one stable compound – silver(I) chloride AgCl (a well-known photosensitive agent in traditional photography). There exists a variety of other compounds containing both silver and chlorine, e.g. ionic conductors Ag$_5$Te$_2$Cl [4] and AgF$_{1-x}$Cl$_x$ [5], complex chlorides such as Ag[AuCl$_4$] [6] and Ag[AlCl$_4$] [7], or even ternary chloroargentates, such as CsAgCl$_2$ [8,9], Cs$_2$AgCl$_3$ [9] and Rb$_2$AgCl$_3$ [9,10]. All of the above, however, feature silver in the oxidation state +1 – as does AgCl. An interesting example can be found in the case of Cs$_2$Ag$^I$Ag$^{III}$Cl$_6$ – a mixed-valence chloroargentate with a perovskite-like structure, in which Ag centers are coordinated by Cl atoms in compressed (for Ag$^I$) or elongated (for Ag$^{III}$) octahedra. [11] The latter compound serves as an example of charge transfer instability of Ag$^{II}$ species in homoleptic environment of ligands less electronegative than fluorine, evidenced also by the case of Ag$^I$Ag$^{III}$O$_3$ [12,13], which has been found to retain its mixed-valent (and insulating) character even when compressed to 80 GPa [14]. This tendency, along with the aforementioned high reactivity of Ag$^{II}$ species (towards ligands less electronegative than fluorine), can be used to explain the scarcity of known AgCl$_x$ compounds. Nevertheless, a possibility of obtaining novel binary silver chlorides has been investigated computationally in recent years. Our previous studies have looked at the relative stability of various structural candidates for AgCl$_2$ at ambient [15] and elevated pressure [16], and a different work has identified a candidate structure for a silver subchloride [17]. All of the considered candidates turn out to be metastable with respect to decomposition into AgCl. However, the list of possible candidates for new AgCl$_x$ compounds is not exhausted by the aforementioned studies. As another example of possible configurations, one can mention recently discovered higher chlorides of sodium [18,19] and potassium

[20]. The ionic radius of Ag(I) is similar to that of Na(I) [21] and thus chemistry of polychloride species could also be similar. Indeed, we argue that the experimental data obtained and reported in this contribution points to formation of a hitherto unknown compound which contains Ag(I) and polychloride $Cl_x^-$ anions.[22]

**Methodology**

High-pressure experiments involving mixtures of Ag or AgCl with $Cl_2$ were performed using a diamond-anvil cell (DAC) supplied by Almax EasyLab, with diamond culets 250 µm in diameter. Gaskets were prepared as follows: a rhenium gasket blank was indented to ca. 10 GPa (using ruby fluorescence as pressure gauge [23]), after which a hole ca. 100 µm in diameter was cut out using an infrared laser. A cylinder-shaped piece of hastelloy, matching in size, was inserted into the hole and welded together with the rhenium gasket through further compression, after which a hole was again cut out. The gasket was placed on one of the seats of the DAC and fit tightly to the diamond anvil using a Seger ring. Gaseous $Cl_2$ was then loaded from a capillary into the gasket hole while the entire system was being cooled in liquid nitrogen inside an argon-filled glove-box with desiccant, ensuring condensation of chlorine uncontaminated by moisture. Ag or AgCl (depending on the experiment) was loaded by placing a microcrystalline sample on the other diamond anvil.

Raman spectra were collected using a 532 nm laser. Pressure in samples was determined using the high-edge frequency of first-order Raman band of diamond from diamond anvil [24]. Infrared laser was used in selected experiments to heat the studied Ag-$Cl_2$ sample.

**Results**

We performed several runs of compression experiments with either Ag or AgCl loaded together with $Cl_2$ into a DAC. In general, assuming no reaction taking place, we expect the Raman spectrum to be dominated by features originating from $Cl_2$ – a molecular crystal (chlorine solidifies at room temperature at pressure as low as 1.45 GPa [25]). AgCl, as an ionic solid with rocksalt structure at ambient pressure, would likely produce only relatively low-intensity overtone bands below 400 cm$^{-1}$ at ambient pressure [26,27]. In addition, AgCl undergoes a series of phase transitions in the range 7-13 GPa, ultimately leading to CsCl-type polymorph [28]. Although Raman spectra of CsCl-type AgCl have not been reported, selection rules indicate that only the overtone of IR-active $T_{1u}$ mode can be Raman-active, as is the case with isostructural AgF at high pressure [29]. We simulated the pressure dependence of those frequencies using DFT calculations for reference in further discussion (methodology and results are shown in Supplemetary Material [SM]). Similarly, Ag, as a metal also with fcc crystal structure, is not expected to produce any Raman signal at all. Accordingly, in one of the experiments – where we compressed AgCl with $Cl_2$ and measured Raman spectra at various pressure points between 8 and 52 GPa (initially without any laser heating) – we observed bands characteristic of solid $Cl_2$, whose frequencies and pressure dependence were previously reported by Johannsen *et al.* [30].

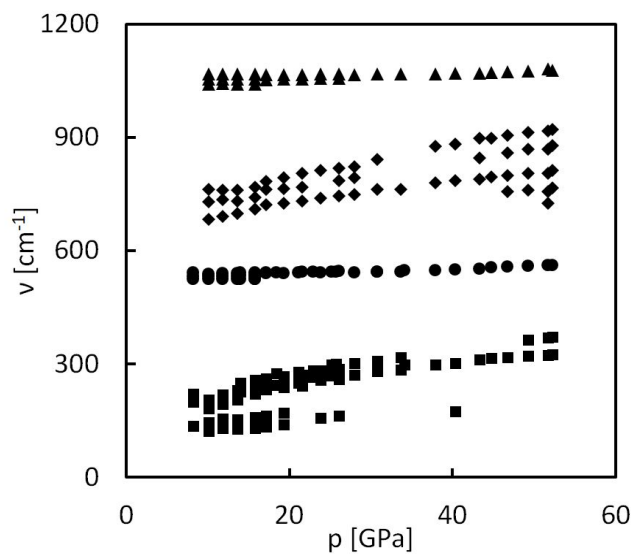

Figure 1. Pressure dependence of bands identified as originating from solid molecular chlorine, taken from one of the experimental runs. Based on ref. [30], we assign these bands as: squares – lattice phonons, circles – intramolecular Cl-Cl vibrations, diamonds – combination modes (Cl-Cl vibration + a lattice phonon), triangles – Cl-Cl vibration overtone. (More detailed information on the assignment can be found in SM.)

In fig. 1 we plot chlorine band frequencies measured in this work as a function of pressure. Overall, our results for chlorine are in very good agreement with previously reported data [30], although we extended our assignment to bands appearing in the region up to ca. 1100 cm$^{-1}$. We present example spectra and more detailed assignment in SM. For the purpose of this work, signals originating from $Cl_2$ serve as a backdrop for identification of additional bands, which were observed in further experiments, as discussed below.

We performed two other experimental runs: (A) AgCl + $Cl_2$ mixture compressed to ca. 40 GPa, laser-heated several times and further compressed to ca. 60 GPa in several steps, and (B) Ag + $Cl_2$ mixture compressed initially to ca. 4 GPa and further compressed to ca. 27 GPa in several steps, without laser heating. Raman spectra were collected at each compression step in both samples.

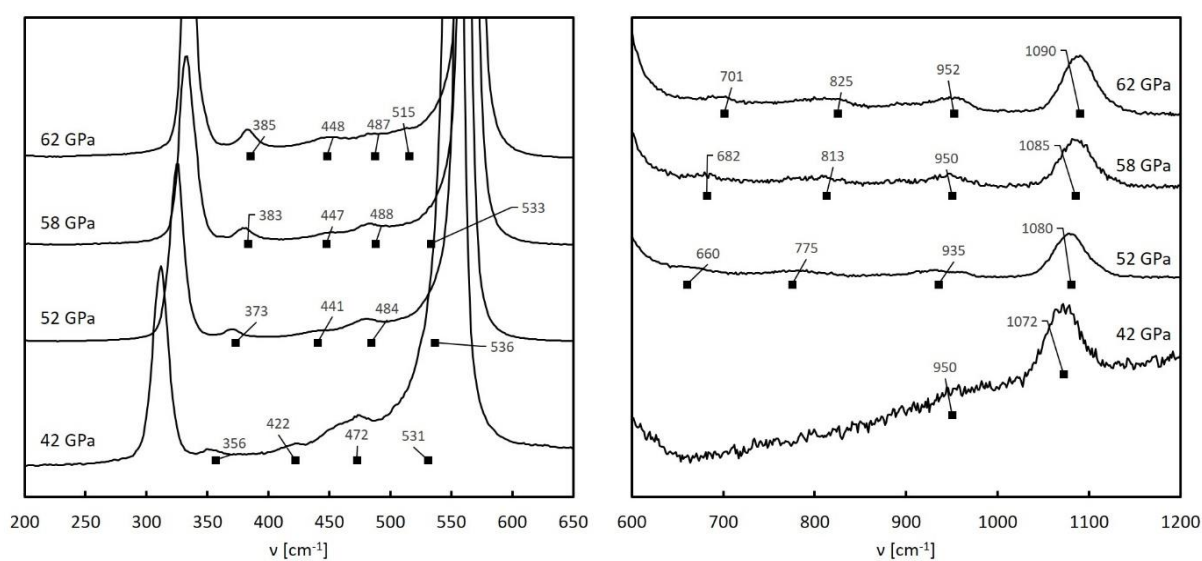

Figure 2. Comparison of spectra obtained in experiment A: AgCl + $Cl_2$ mixture compressed initially to ca. 40 GPa. See text for further details.

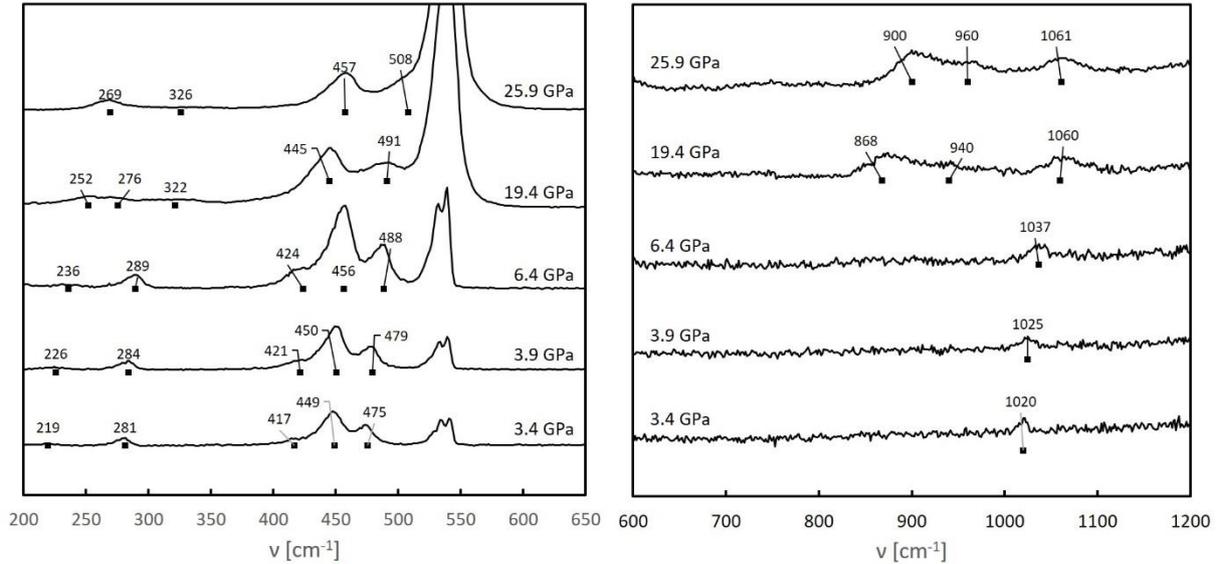

Figure 3. Comparison of spectra obtained in experiment B. Ag + $Cl_2$ mixture compressed initially to ca. 3 GPa. See text for further details.

In fig. 2, we present the spectra collected for sample A. The two most prominent ones can be assigned to chlorine as Cl-Cl vibron (at ca. 550 cm$^{-1}$ at 42 GPa) and as $A_g$ lattice phonon (at ca. 310 cm$^{-1}$ at 42 GPa). The smaller signal in the range 350-390 cm$^{-1}$ (depending on pressure) is most likely the $B_{3g}$ solid $Cl_2$ lattice phonon. In the higher-frequency region, the previously mentioned overtone of Cl-Cl vibron is also visible at 1070-1090 cm$^{-1}$. Importantly, we can discern several new bands appearing in the range just below the Cl-Cl vibron (400-500 cm$^{-1}$). The bands are relatively weak, but repeatable, and the spectrum is still dominated by features from $Cl_2$. Several weak signals can also be discerned in the higher-frequency region (600-1000 cm$^{-1}$). It is important to stress that these new bands appeared after laser-heating and therefore cannot be assigned to AgCl, which was present in the anvil from the beginning. Overtone bands which could originate from CsCl-type AgCl (*cf*. SM) are not observed in the spectrum, indicating very low Raman activity of this system.

Raman spectra collected from experiment B are shown in fig. 3. In this case we can also observe new bands appearing in the 400-500 cm$^{-1}$ region. Interestingly, this occurs in sample B even without laser heating. Assuming that the origin of these new signals is a product of reaction in the Ag-$Cl_2$ system, such observation is not surprising – we can expect that the activation barrier for the reaction between metallic Ag and $Cl_2$ will be lower than for the respective reaction of AgCl and $Cl_2$, since AgCl is thermodynamically stable with respect to Ag and $Cl_2$ at room temperature within the studied pressure range [16,28]. The possible origin of new bands appearing in figs. 2 and 3 is discussed below.

The appearance of bands which cannot be assigned to any of the constituents of the initial mixture clearly indicate that a reaction has taken place. The system Ag/AgCl + $Cl_2$ can in principle also interact with the gasket material (hastelloy, made up mostly of Ni, Cr and Mo) and the diamond anvil, particularly in the heated samples [31]. However, we have chosen hastelloy as a gasket material due to its chemical resistance, and the reaction between it and $Cl_2$ (especially without laser heating) appears unlikely. Known spectral data for possible products of such reaction, i.e. nickel, chromium and molybdenum chlorides, also testify against such assignment of the new bands in 400-500 cm$^{-1}$ region [32–34]. Furthermore, a reaction between $Cl_2$ and diamond could conceivably occur, although previous experimental works with similar setup and heating do not mention such outcome [35]. Chlorinated diamond surfaces have been analyzed by Raman spectroscopy [36]; C-Cl vibrations in such systems are expected to fall in the 600-800 cm$^{-1}$ range. In this work, a faint band appearing in sample A between

660 and 700 cm$^{-1}$ may originate from such vibrations. Its intensity is very weak and thus the extent of such chlorination can be considered very low, if it occurs at all. Finally, CCl$_4$ could in principle also be a product of reaction between diamond and Cl$_2$. Raman spectra and phase transitions of CCl$_4$ have been previously studied, and known frequencies of its bands from those works have been taken into account when analyzing new bands in experiments A and B [37].

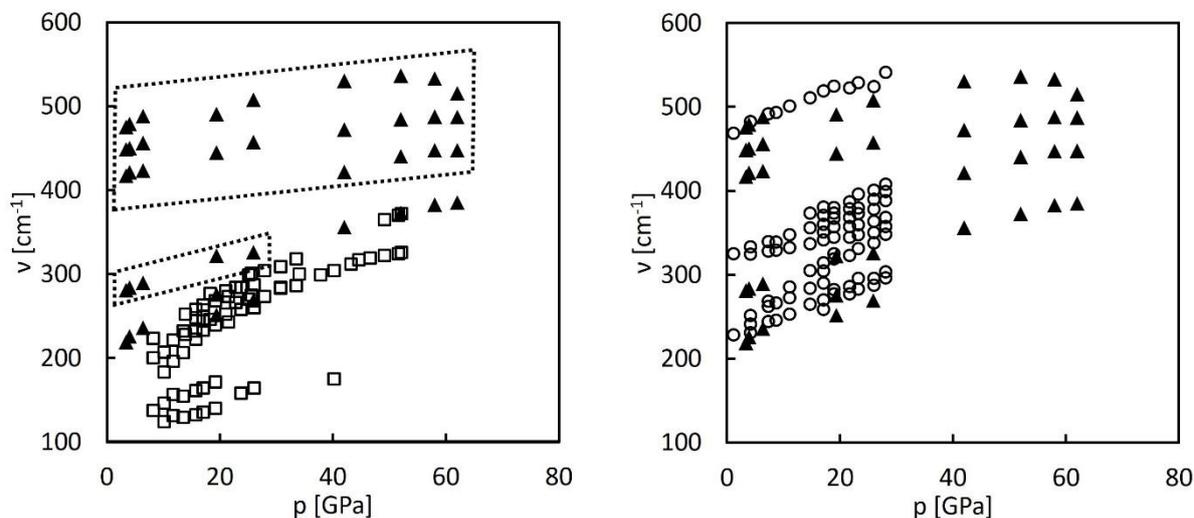

Figure 4. Comparison of new bands appearing in spectra from experiments A and B (solid triangles) with: left panel – known bands from solid chlorine (this work, hollow squares), right panel – known bands from solid CCl$_4$ (ref. [37], hollow circles). Cl-Cl intramolecular vibron band appearing at 550-600 cm$^{-1}$ is omitted for clarity. Dashed parallelograms indicate bands unaccounted for by Cl$_2$ or CCl$_4$.

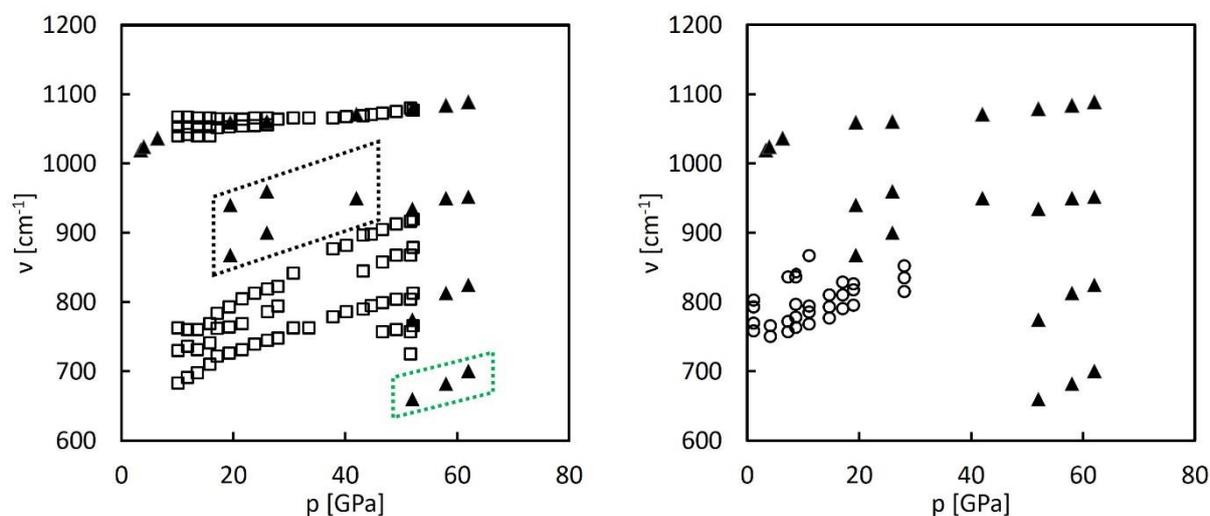

Figure 5. Comparison of new bands appearing in spectra from experiments A and B (solid triangles) with: left panel – known bands from solid chlorine (this work, hollow squares), right panel – known bands from solid CCl$_4$ (ref. [37], hollow circles). Black dashed parallelogram indicates bands unaccounted for by Cl$_2$ or CCl$_4$. Green dashed parallelogram indicates bands possibly originating from chlorinated diamond.

Figs. 4 and 5 plot pressure dependence of new bands observed in samples A and B, together with previously reported bands of CCl$_4$ [37]. Although the range of data for CCl$_4$ is only up to ca. 30 GPa, we can still see that the new bands in 400-500 cm$^{-1}$ cannot be accounted for by this compound. Several of the higher-frequency bands (in 600-1200 cm$^{-1}$) can be assigned to Cl$_2$ vibron overtones and combination modes mentioned previously. The bands in the 850-950 cm$^{-1}$ region, regardless of their chemical origin, are likely overtones of the bands between 400 and 500 cm$^{-1}$, since in general, overtones are a common occurrence in Raman spectroscopy (*cf.* SM for assignment of overtones and combination modes in $Cl_2$ spectrum). Overtones tend to be strong when Raman spectra have resonance character, which is quite likely in the case of dark brown sample (*cf.* SM) illuminated with visible laser beam.

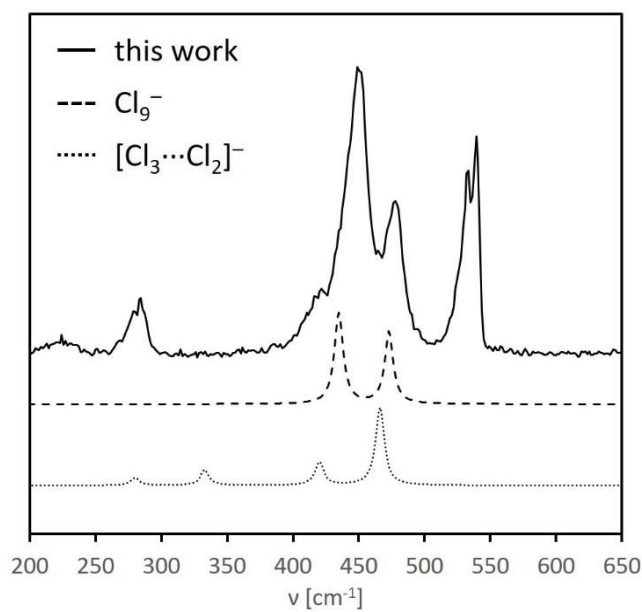

Figure 6. Comparison of the 200-600 cm$^{-1}$ region of sample B spectra at 3.9 GPa with Raman bands of polychloride anions derived from ref. [38] for $Cl_9^-$ anion and from ref. [39] for 2D polychloride network. Relative intensities are also adapted from cited works.

Having ruled out possible candidates for undesirable reaction products, we can attempt to determine the identity of system constituents which give rise to the new bands in the 400-500 cm$^{-1}$ region. Discontinuities in pressure dependence and relative positions of these bands (fig. 4 and 5) between sets A and B suggest that they may belong either to different polymorphs of the same compound (obtained as one phase in experiment A and as another in B) or to different compounds altogether. These bands are located just below the dominant Cl-Cl vibron band, and simple considerations based on the harmonic oscillator model indicate that they could originate from: (a) X-Cl vibrations, where X is an atom heavier that Cl, (b) Cl-Cl vibrations from a species featuring Cl-Cl bonds longer than in $Cl_2$ molecule, such as polychloride anions [38–40], or (c) a lattice phonon in a solid compounds featuring (a) and/or (b). Fig. 6 shows the spectrum obtained for sample B at 3.9 GPa compared with simulated spectra of example polychloride species at ambient pressure from ref. [38] and [39]. While not an exact match, it is worth noting that (i) these values were observed/calculated for different compounds containing complex organic cations, so we can expect these frequencies to vary to some extent depending on the system, and (ii) there is no data on the pressure dependence of Raman spectra of polychlorides, so a certain discrepancy (a positive shift) can be expected between previously reported values and spectra presented in this work. However, the comparison in fig. 6 gives strong credibility to the hypothesis that polychloride species such as $Cl_5^-$ or $Cl_9^-$ are indeed present in the sample. Lastly, the sample obtained in experiment A turned visibly brown at the irradiation spot (*cf.* SM, fig. S6), which indicates formation of a product with a charge-transfer transition in the visible range. Given that Ag(I) polychloride(s) is/are present in the sample, the transition responsible for the color may correspond to metal-to-ligand charge transfer excitation from occupied d states in Ag$^I$ and empty σ* antibonding states in $Cl_2$ molecule within polychloride anion or array (Ag$^I$ d → $Cl_2$ σ*).

**Conclusions**

We have obtained and analyzed Raman spectra probing several experiments with binary Ag-$Cl_2$ system, and we argue that the data is indicative of the formation of an as-yet unknown phase or phases. Based on the positions of new Raman bands observed in the experiments, these new phase(s) most likely contain polychloride anions [40] or in general, a more complex arrangement of Cl atoms [39]. Recent computational results work investigating different stoichiometries of $AgCl_x$ compounds using evolutionary algorithm has indeed pointed to a possibility of formation of polymorphs with polychloride anions for $x > 2$ [41]. A common feature emerging in those solutions are $Cl_3^-$ anions and $[Cl_3…Cl_2]^-$ and $[Cl_2…Cl_2]^-$ infinite networks. Further study is needed in order to elucidate the exact nature of reaction products reported here. Unfortunately, high-pressure synchrotron X-ray diffraction measurements, which could provide insight into the crystal structure of these systems, have been severely delayed due to the ongoing COVID-19 pandemic.

**Acknowledgments**

This work was financed by Polish National Science Center (NCN) through a Preludium grant (2017/25/N/ST5/01976). A.G. would like to thank Dr. Przemysław Malinowski (Center of New Technologies, UW) and Dr. Dominik Kurzydłowski (Faculty of Mathematics and Natural Sciences, UKSW) for their help in preparing and carrying out the experiments. M.D. acknowledges the ERDF, Research and Innovation Operational Programme (ITMS2014+: 313011W085), the Slovak Research and Development Agency (APVV-18-0168) and Scientific Grant Agency of the Slovak Republic (VG 1/0223/19).

# Observation of a new polyhalide phase in Ag-Cl$_2$ system at high pressure

Adam Grzelak[1]*, Jakub Gawraczyński[1], Mariana Derzsi[2], Viktor Struzhkin[3], Maddury Somayazulu[3], Wojciech Grochala[1]

[1]*Center of New Technologies, University of Warsaw, Zwirki i Wigury 93, 02089 Warsaw, Poland*
[2]*Advanced Technologies Research Institute, Faculty of Materials Science and Technology in Trnava, Slovak University of Technology in Bratislava, J. Bottu 8857/25, 917 24 Trnava, Slovakia*
[3]*Center for High Pressure Science and Technology Advanced Research, Shanghai 201203, China*
[4]*HPCAT, X-ray Science Division, Argonne National Laboratory, Lemont, IL 06439, United States*

*\*a.grzelak@cent.uw.edu.pl*

## SUPPLEMENTARY MATERIAL

S1. Raman spectra of solid Cl$_2$ and pressure dependence of band frequencies

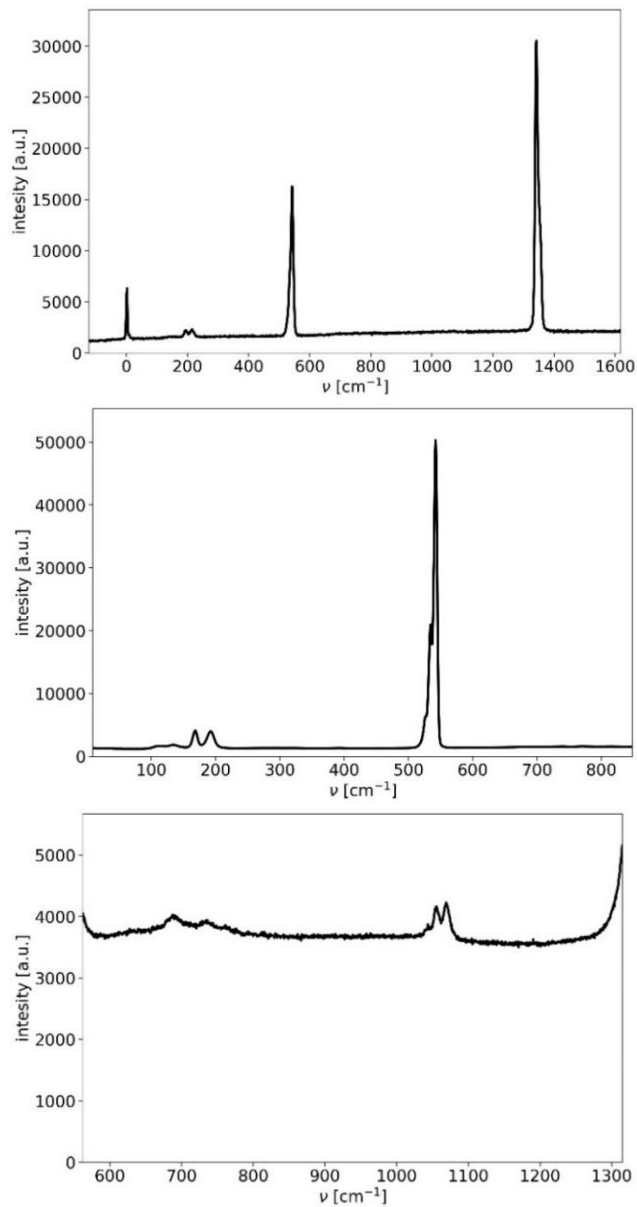

Fig. S1. Spectra of solid chlorine obtained at 8 GPa. The high-intensity band at ca. 1350 cm$^{-1}$ originated from diamond anvil.

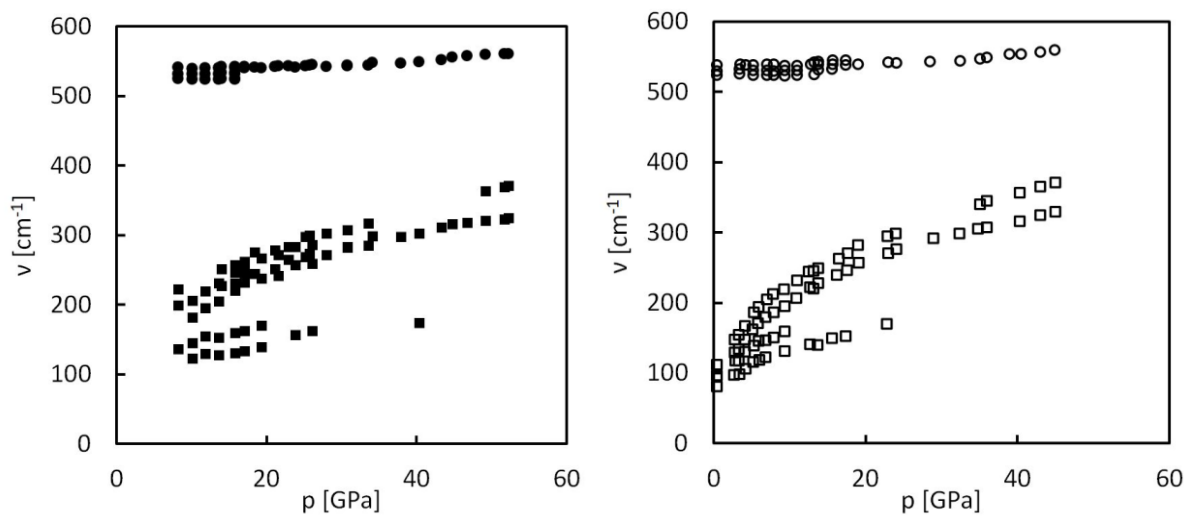

Figure S2. Comparison of pressure dependence of chlorine bands in this work (left panel) and in Johannsen *et al.* [1]. Circles are Cl-Cl$_2$ intramolecular vibrations and squares are lattice phonons.

Raman spectra for solid Cl$_2$ obtained in this work, and subsequent pressure dependence of bands identified in those spectra, are in good agreement with ref. [1] (fig. S2). Accordingly, the bands appearing in the region ca. 100-400 cm$^{-1}$ are designated as lattice modes – from lowest to highest frequency: $B_{1g}$, $B_{2g}$, $A_g$, $B_{3g}$. Disappearance of $B_{1g}$ and $B_{2g}$ modes above certain pressures is also in agreement with ref. [1]. The bands at ca. 520-550 cm$^{-1}$ originate from intramolecular Cl-Cl vibrations. The main band in this group is also the most intense in the spectrum of chlorine. Its splitting, i.e. shoulders appearing at lower pressures, is an isotopic effect [1]. Pressure dependence of this band is relatively flat, testifying to rigidity of the Cl-Cl bond, as compared to weak intermolecular interactions in the Cl$_2$ molecular crystal.

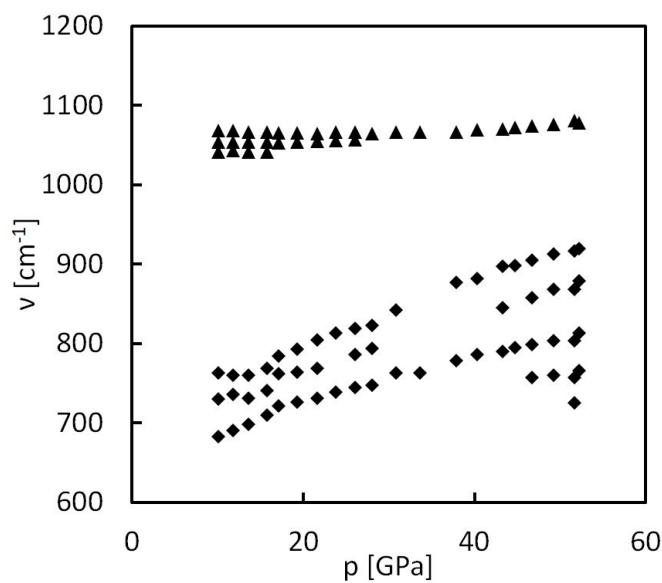

Figure S3. Pressure dependence of bands designated as originating from Cl$_2$ in the 600-1200 cm$^{-1}$ range – this work.

In fig. S3 we plot the pressure dependence of bands in the region above 600 cm$^{-1}$, which were not discussed or shown in ref. [1]. The bands appearing at ca. 1020-1080 cm$^{-1}$ are interpreted to be overtones of intramolecular Cl-Cl vibration bands mentioned above, due to their frequency, slope of pressure dependence, and splitting. The bands appearing in the ca. 650-900 cm$^{-1}$ range are most likely combination bands – Cl-Cl vibron ($A_g$) + one of the lattice phonon modes ($B_{1g}$, $B_{2g}$, $B_{3g}$). These bands are very weak in intensity compared to the rest of those discussed. Multiplying the corresponding irreducible representations, we get the overtones: $B_{1g} \times A_g = B_{1g}$, $B_{2g} \times A_g = B_{2g}$, and $B_{3g} \times A_g = B_{3g}$. An example analysis of such combination mode is shown in fig. S4.

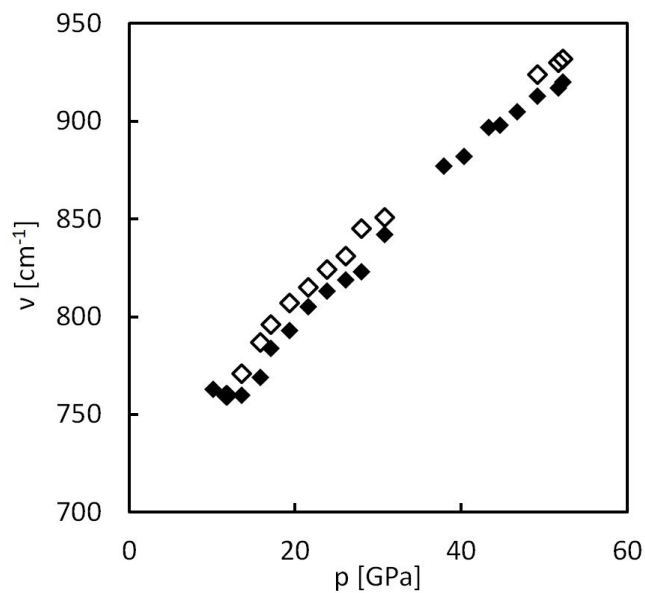

Figure S4. Pressure dependence of one of the combination modes ($B_{3g} \times A_g = B_{3g}$) observed in the spectrum of $Cl_2$. Filled diamonds – observed frequencies, hollow diamonds – pressure dependence of the sum of frequencies of $B_{3g}$ lattice phonon and $A_g$ $Cl_2$ vibron. Note the anharmonicity of the observed combination mode, i.e. a lower observed frequency compared to the simple sum.

## S2. Pressure dependence of simulated (DFT) frequencies of the overtone of IR-active $T_{1u}$ mode of AgCl in CsCl structure.

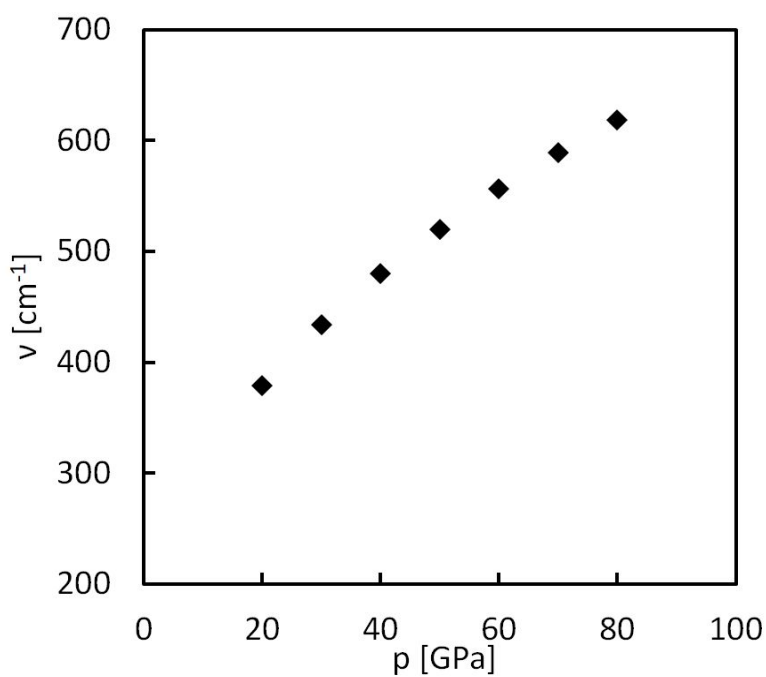

Figure S5. Pressure dependence of simulated (DFT) frequencies of the overtone of IR-active $T_{1u}$ mode of AgCl in CsCl structure.

Fig. S5 shows pressure dependence of simulated frequency of the overtone of $T_{1u}$ mode in CsCl-type AgCl ($Pm\overline{3}m$), which is a stable polymorph of AgCl above ca. 13 GPa [2]. Calculations were carried out using VASP software [3–7], with GGA-type Perdew-Burke-Ernzerhof functional adapted for solids (PBEsol) [8] was used. Plane-wave cutoff energy was set to 800 eV. Integration grid of 11x11x11 k-points was used.

S3. Picture of the one of the samples

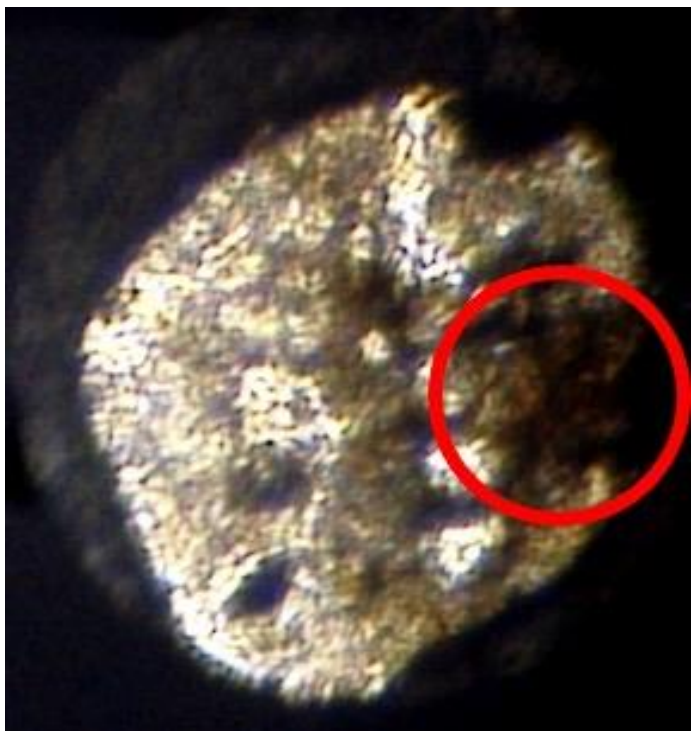

Figure S6. Picture of sample from experiment A after initial compression and laser heating. Note the brown color of the irradiated spot.